\documentclass[reprint]{revtex4-1}
\usepackage{graphicx}%
\usepackage{dcolumn}%
\usepackage{bm}%
\usepackage{upgreek}

\usepackage[colorlinks=true,linkcolor=blue]{hyperref}%
\expandafter\ifx\csname package@font\endcsname\relax\else
 \expandafter\expandafter
 \expandafter\usepackage
 \expandafter\expandafter
 \expandafter{\csname package@font\endcsname}%
\fi
\begin{document}

\title{100GHz Integrated All-Optical Switch Enabled by ALD}%

\author{Gregory Moille, Sylvain Combri\'{e}, Ga\"{e}lle Lehoucq, Bowen Hu, and Alfredo De Rossi}
\affiliation{Thales Research and Technology, 1 Avenue A. Fresnel, 91767
Palaiseau, France}
\author{Laurence Morgenroth, Fran\c{c}ois Neuilly and Didier Decoster}
\affiliation{Institut d'Electronique et de Micro et Nanotechnologies, Villeneuve d'Asq, France}

\date{\today}%

\begin{abstract}
The carrier lifetime of a photonic crystal all-optical switch is optimized by controlling the surface of GaAs by Atomic Layer Deposition. We demonstrate an all optical modulation capability up to 100GHz at Telecom wavelengths, with a contrast as high as 7dB. Wavelength conversion has also been demonstrated at a repetition rate of 2.5GHz with average pump power of about 0.5mW
\end{abstract}

\maketitle

\section{Introduction}
\begin{figure}[b]
	\begin{center}
	\includegraphics[width=0.8\columnwidth]{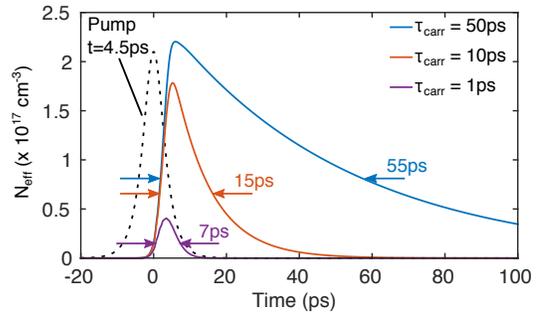}
	\caption{\label{fig:CMT_N_vs_tcarreps} Calculated density of free carriers excited inside cavity excited by a 4.5ps pulse (black dashed line), depending on their lifetime $\tau_{carr}$. Arrows represent the gating window at $1/e$.}
	\end{center}
\end{figure}

Complex photonic circuits, particularly optical interconnects, would greatly benefit from avoiding the systematic conversion into and from the electrical domain by using fast optically-controlled optical switches.
In this respect, three requirements are critical: the power consumption, the speed and the footprint. On top of that is the inherent compatibility with a photonic integrated platform. All-optical signal processing requires a nonlinear response which, unfortunately, is either weak, or slow in most accessible materials; nevertheless, a standard single mode optical fiber can be configured to operate as a very fast and efficient all-optical gate by exploiting a long interaction length.\cite{doran_nonlinear-optical_1988} This device enables ultra-fast signals at THz rate\cite{mulvad_demonstration_2010} but is incompatible with photonic integrated circuits.\\
Resonant enhancement in compact semiconductor integrated circuits has been successfully employed to demonstrate all-optical signal modulation \cite{absil_wavelength_2000,almeida_all-optical_2004}. Here, the amplitude modulation results from the spectral shift of the resonance. The related refractive index change follows the generation of a plasma of free carriers by an optical pump and it is orders of magnitude stronger than the Kerr effect in a silica fiber. Thus, the combination of a free carrier-related index change and the very tight confinement of the optical field in photonic crystal (PhC) cavities was preferred for energy-efficient and fast integrated all-optical switching\cite{tanabe_all-optical_2005,husko_ultrafast_2009,combrie_all-optical_2013,heuck_heterodyne_2013,bazin_ultrafast_2014,yu_nonlinear_2014, yu_nonreciprocal_2015} ultimately leading to ultra-efficient operation at the femto-Joule level,\cite{nozaki_sub-femtojoule_2010}.\\
In a free-carrier based switch, the density of the carrier population $N$ in the cavity, to which the change in the refractive index $\Delta n$ is proportional, builds up depending on the simple rate equation $\partial_t{N}=-N/\tau_{carr}+G(t)$,  and, therefore, the carrier lifetime $\tau_{carr}$ should be ideally close to the duration of the excitation $G(t)$, otherwise the response is either too weak (carriers have no time to build up) or too slow. This is apparent in Fig.\ref{fig:CMT_N_vs_tcarreps}, where we consider the excitation of free carriers into a nonlinear resonator with a pump pulse with duration ($t_{pulse}$=4.5 ps) compatible with 100 GHz data processing. Particularly, there is little gain in the maximum achievable density when $\tau_{carr}$ substantially exceeds $t_{pulse}$, while the penalty in bandwidth is drastic. In an optimized all-optical switch, the carrier relaxation time should be adjusted relative to the desired operating speed.\\
Measurements on PhC cavities made of Silicon\cite{tanabe_all-optical_2005,tanabe_carrier_2008} and Indium Phosphide\cite{nozaki_sub-femtojoule_2010,heuck_heterodyne_2013} reveal a nonlinear response with a fast time constant of about 10 to 20 picoseconds, and a slower contribution with a time constant of a few hundreds of picoseconds. The recovery time of the gate is related to the carrier dynamics through a function which depends on the line-shape of the resonator and on the relative detuning of the signal to be modulated. A suitable optimization of this function may result in a faster device, \cite{yu_ultrafast_2015} however it seems difficult to exceed a pulse rate of 20 GHz without avoiding unrecoverable interference between adjacent symbols. The modification of the semiconductor structure by addition of a InGaAs surface layer to a InP PhC resulted into a single-exponential carrier relaxation time constant of about 30 ps, a clean gate resolution of about 20 ps and all-optical signal processing at 10 GHz\cite{bazin_ultrafast_2014}. Interestingly, pump-probe measurements on Silicon micro-ring resonators\cite{Waldow2008} and PhCs\cite{Tanabe2007} have shown a decrease of the recovery time down to 25 ps, owing to a ion implantation.

\noindent A much faster dynamics was reported in GaAs photonic crystals\cite{bristow_ultrafast_2003,husko_ultrafast_2009} using pump-probe techniques, suggesting that the carrier relaxation time is in the picosecond range, however, the modulation contrast was very low and it has not been possible to achieve wavelength conversion.\\
 In this manuscript, we demonstrate the optimization of the carrier relaxation time in GaAs PhC switch to about 10 ps, which is optimal for ultra-fast operation. This was accomplished by passivating the surface of  using a $Al_2O_3$ conformal coating by Atomic Layer Deposition (ALD), which we show that effectively reduces the surface recombination velocity by more than an order of magnitude. This device could process data at a 100 GHz rate and still meets the requirements of footprint,  power consumption and compatibility with a photonic integrated circuit.
\section{device description}
\begin{figure}[b]
	\begin{center}
	\includegraphics[width=\columnwidth]{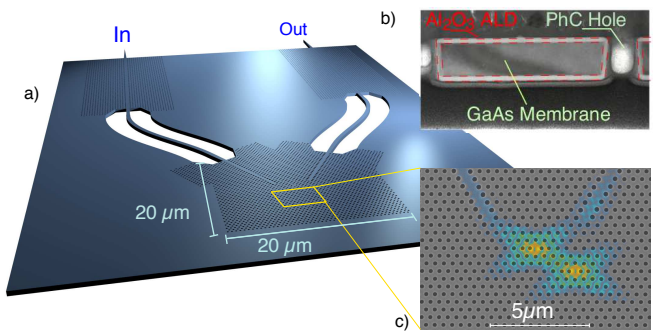}
		\caption{\label{fig:Sample}a) Artist's view of the sample, b) TEM cross section of the PhC slab revealing the conformal $Al_2O_3$ coating and c) SEM image (top view) of the device core, the two coupled cavities with the electric field intensity map superimposed (logarithmic scale).}
	\end{center}
\end{figure}

\begin{figure}[b]
	\includegraphics[width=0.95\columnwidth]{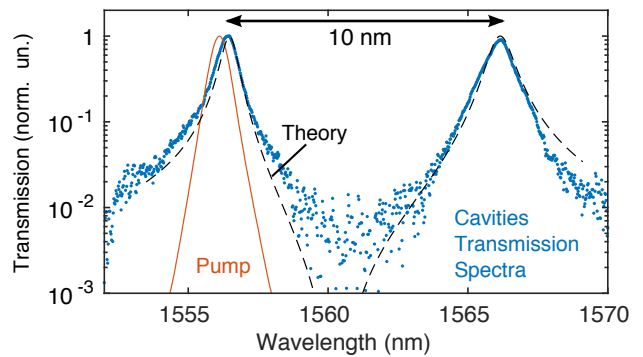}
	\caption{\label{fig:linear} Transmission ($P_{out}/P_{in}$) spectrum, measured through the sample. The two resonances are located  at $\lambda_b=1557$nm and $\lambda_r=1567$nm with Q-factor $\approx 3000$. The theoretical lineshape (Coupled Mode Theory) is superimposed (black dashed line). The spectrum of the pump with offset $\Delta\lambda_{p}$ is depicted (red line)}
\end{figure}

The nonlinear resonator (Fig.\ref{fig:Sample})  consists of two coupled, nominally identical $H_0$ photonic crystal cavities, forming a \textit{photonic molecule}, following the same design as in ref. \cite{combrie_all-optical_2013}. Details are given in the methods section. They are connected to the input and the output channels using a short section of suspended photonic wire, PhC waveguides and a mode adapter\cite{tran2009apl} for coupling to the optical lensed fiber. The footprint of the functional part of the device (excluding coupling) is about $20 \times 20 \upmu \mathrm{m}$.
The coupling strength  results into a spectral separation of the two resonances by about 10 nm (Fig.\ref{fig:linear}). Interestingly, the transmission spectra are remarkably smooth and fit very well with the ideal lineshape predicted by the Coupled Mode Theory (CMT). This indicates that parasitic reflections, scattering and disorder effects are negligible.
The two resonance peaks reach exactly the same transmission level as expected from the design, while the transmission minimum is 30 dB below.\\
\noindent The fabrication of the sample is completed using the ALD technique, which ensures a conform growth over any exposed surface of the PhC structure. More common deposition techniques, used to passivate the surfaces of semiconductors\cite{yablonovitch_nearly_1987}, do not ensure a perfect encapsulation of complicated structures such as air-suspended PhC slab, which is here critical. ALD has been used to passivate GaAs nanowires\cite{Herrmann2005Appl.Phys.Lett.} or to protect photonic crystal from photo-induced oxidation\cite{Kiravittaya2011J.Appl.Phys.}.
In our sample, the oxide is removed from the surface by \textit{in situ} cleaning, then about 30 nm of Al$_2$O$_3$ are deposited (Fig.\ref{fig:Sample}c). We have chosen Al$_2$O$_3$ because it induces a high density of negative charges at the interface\cite{KotipalliR.2013EPJPhotovolt}. In GaAs, this negative charge may be enough to unpin the Fermi level from mid-gap\cite{Robertson2009Microelectron.Eng.}, which results into a drastic decrease of the surface recombination. ALD has been successfully used to increase the carrier lifetime in solar cells\cite{Mariani2013NatCommun}.
%

\section{Dynamical response}
In the conditions of operation, the device dynamics is dominated by the excitation, diffusion and recombination of the free-carriers and it is captured by a rate equation model coupled to time-dependent coupled mode theory\cite{manolatou_coupling_1999}. The device response depends on the instantaneous detuning of each cavity $\Delta\omega_l$, which, in turns, depends on the effective carrier density obtained by weighting the distribution of the carriers $N$ with the normalized field $u$, i.e. $N_{eff}=\frac{1}{2}\epsilon_0\int \epsilon_r(\mathbf{r})  N(\mathbf{r})|u|^2 \mathrm{d}V$.  In the ambipolar approximation, which holds here, the distributions of electrons and holes coincide and induce a local change in the index of refraction $\frac{\partial n}{\partial N}$ entailing three contributions from bandgap shrinkage, band filling and free-carrier dispersion, each of them strongly dispersive such that their relative weight depends on the spectral range considered\cite{bennett_carrier-induced_1990}. Thus, the instantaneous cavity detuning is simply related to the effective carrier density at each cavity: $\Delta\omega_l=\frac{\partial\omega}{\partial n}\frac{\partial n}{\partial N}N_{eff}^{(l)}$.\\

\begin{figure}[t]
	\includegraphics[width=\columnwidth]{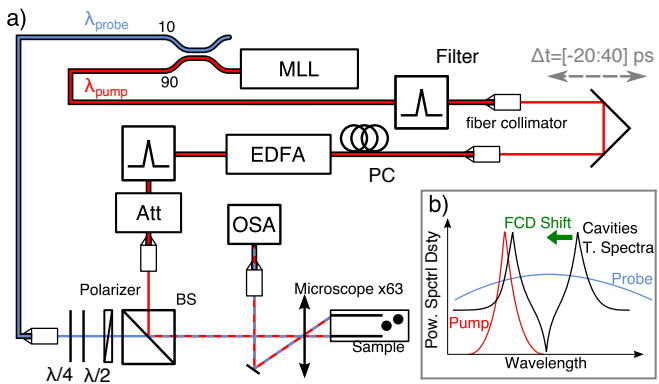}
	\caption{\label{fig:setup}a) Experimental set-up for time-resolved spectral measurements. OSA: Optical Spectral Analyser, BS: Beam Splitter, EDFA: Erbium Doped Fiber Amplifier, Att: Attenuator, PC: Polarizing Control, MLL: Mode-Locked Laser. b) Schematic in the spectral domain: blue: broad probe spectrum, red: pump spectrum, black: cavity transmission spectra. }
\end{figure}

It has been shown\cite{bristow_ultrafast_2003,raineri_ultrafast_2004} in nonlinear photonic crystals that the instantaneous detuning can be retrieved by time-resolved spectral analysis. The technique consists in recording the transmission (or the reflection) spectra of a short probe pulse used as a \textit{white light} source, as a function of its delay relative to the pump. A carrier relaxation constant of about 8 ps was thus estimated from the time evolution of the photonic band-gap of a AlGaAs photonic crystal structure\cite{bristow_ultrafast_2003}. It was pointed out that the surface recombination at the holes of the PhC is responsible for about one order of magnitude decrease of $\tau_{carr}$. A cavity recovery time of about 6ps was observed in a GaAs PhC membrane resonator by homodyne pump-probe analysis\cite{husko_ultrafast_2009}, which suggests a much faster carrier relaxation time, once accounting for the duration of the excitation (about 4 ps)  and the cavity lifetime (1 ps). Such a short lifetime is consistent with the modelling of the carrier diffusion within the PhC, namely $\tau_{carr}\approx 1\,\mathrm{ps}$ , assuming a surface recombination velocity $S$ ranging from $3\times 10^4$ to $10^5\, \mathrm{m.s}^{-1}$, corresponding to non passivated GaAs surface\cite{nolte_surface_1990,cadiz_surface_2013}. Interestingly, the Shockley formula\cite{shockley1950electrons} $\tau^{-1}_{carr}=\tau^{-1}_{rec}+L^{-1}S \approx L^{-1}S$ gives a good estimate $\tau_{carr}\approx$ 0.8 to 3ps, when $L$ is interpreted as the maximal distance from the surface, which is here about 100 nm.\\
Such as short carrier lifetime explains the small modulation contrast (1dB) observed and may therefore not be desirable when operating with pulses with duration of several picoseconds\cite{husko_ultrafast_2009}.
Furthermore, it has been shown that non passivated GaAs photonic crystal undergoes to a photo-assisted oxidation, resulting into an irreversible drift of the resonance\cite{lee_local_2009}. An irreversible and strong drift of the resonance was also observed when exciting the GaAs nonlinear cavity investigated in ref. \cite{husko_ultrafast_2009} with a GHz repetition rate.\\

The recovery dynamics of the all-optical gates as in Fig.\ref{fig:Sample} is measured with a time-resolved spectral technique similar to\cite{bristow_ultrafast_2003,raineri_ultrafast_2004}, except that here pump and probe are both in the Telecom spectral range, as in ref.\cite{husko_ultrafast_2009}, and are obtained from a femto-second fiber laser (Optisiv). Since GaAs is transparent, free carriers are generated by resonant two-photon  absorption (TPA) in the cavity only. The probe is approximately 80 nm broad and almost flat, and the pulse duration is approximately 1 ps. The pump is temporally shaped and tuned using an optical bandpass filter, while  the relative pump-probe delay is controlled mechanically with a translation stage (Fig. \ref{fig:setup}). The spectral power density of the probe is at least 2 orders of magnitude below the pump. Moreover, the same measurement was repeated by varying the probe level to confirm the absence of any influence on the dynamics.\\
\begin{figure}[tb]
	\begin{center}
	\includegraphics[width=.95\columnwidth]{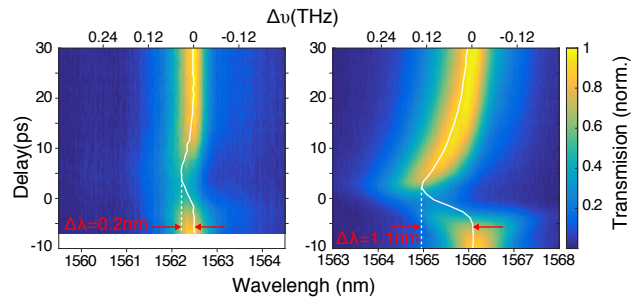}
	\caption{\label{fig:Expe_Map} Time-resolved transmission spectra (normalized intensity maps, linear scale). a) Reference b) Al$_2$O$_3$-coated.  The first order momentum $\nu_c(t)$ is superimposed (white solid line)}
	\end{center}
\end{figure}

\begin{figure}[t]
	\begin{center}
	\includegraphics[width=.95\columnwidth]{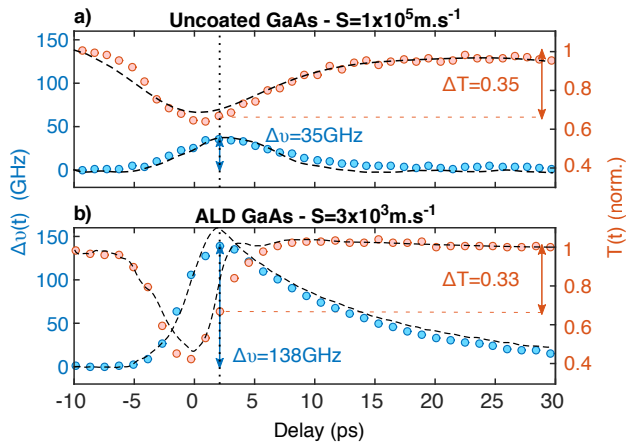}
	\caption{\label{fig:spectral_shift}  Instantaneous detuning $\Delta\nu(t)=\nu_c(t)-\nu_c(-\infty)$, left scale and time-resolved transmission $T(t)$, extracted from Fig.\ref{fig:Expe_Map}.  Reference a) and ALD passivated b) and  devices. The (maximum) pump energy is estimated to 300 fJ. The solid dashed line represents the modelled experiment. Calculations in a) and b) only differ by the value of surface recombination velocity $S$.}
	\end{center}
\end{figure}

\noindent The operation of this nonlinear photon molecule is understood by considering that the field distributions corresponding to the two resonances overlaps almost perfectly (Fig. \ref{fig:Sample}c). Therefore, the two cavities are excited by the same amount of energy when a $sech^2$ pump pulse with duration of approximately 4.5 ps is spectrally matched (Fig. \ref{fig:linear}) to one of the two resonances (with possibly a small detuning $\Delta\lambda_p$). In a highly simplified picture, the two resonances would experience the same spectral shift, which is indeed confirmed by modelling (Fig. \ref{fig:spectral_shift}). Thus, the measured time-resolved spectral shift of the other resonance is directly related to the effective carrier density in the cavities.\\
The time-resolved spectral measurements (Figs. \ref{fig:Expe_Map} and \ref{fig:spectral_shift}) reveal a striking difference between the ALD-coated and the reference (uncoated) sample, (Fig.\ref{fig:Expe_Map}), which is particularly apparent when considering the spectral shift. A quantitative measure is the first order momentum of the time-resolved transmitted optical spectra $T(\nu,t)$, namely: $\nu_c(t)=T(t)^{-1}\int{\nu T(\nu,t)d\nu}$, with $T(t)=\int{T(\nu)d\nu}$ the time-resolved transmission.\\
\noindent  Let us first consider the un-passivated sample. It is apparent from Fig.\ref{fig:spectral_shift}a, that the spectral shift $\Delta\nu$ is fairly weak (35 GHz) when compared to the case of the ALD-coated sample (138 GHz). Also, the temporal shape of $\Delta\nu(t)$ merely follows the instantaneous transmission $T(t)$, suggesting that the device is responding faster than the excitation ($t_{pulse}=4.5$ ps). We performed modelling by including the transmission of the broad probe and compared the measured and calculated $\Delta\nu(t)$ and $T(t)$, also shown in Fig.\ref{fig:spectral_shift}. Importantly, the parameters for the linear CMT equations are obtained by fitting the transmission spectra (Fig. \ref{fig:linear}), while the excitation pulse is entirely determined by the spectral and autocorrelation measurements and the nonlinear interaction terms are calculated. Thus, the only adjustable parameter is the surface recombination velocity $S$. We obtain an excellent agreement on both $T$ and $\Delta\nu$ when $S=10^5\mathrm{m.s}^{-1}$  and $S=3\times10^3\mathrm{m.s}^{-1}$  for the uncoated and the coated sample respectively.
When inspecting the calculated decay of the carrier density, this translates into carrier lifetimes respectively equal to 1.1 ps and 12 ps, which is consistent with a direct analysis of  $\Delta\nu$, as discussed earlier. Moreover, the exponential fit of the trailing edge of $\Delta\nu$ relative to the ALD-coated sample yields about 10 ps, indicating that the relaxation of the frequency shift is following the decaying carrier population after the excitation has extinguished.
The analysis of Fig.\ref{fig:spectral_shift}b near zero delay deserves some further comments.  The transmission minimum $T=0.4$ is an unavoidable consequence of the fact that carriers are injected through nonlinear absorption. This minimum coincides with the steepest slope of the leading edge of $\Delta\nu$, which itself reaches a maximum when the transmission has almost recovered. This behaviour is crucial for a practical use of the switch and it is in striking contrast with the case in Fig.\ref{fig:spectral_shift}b. \\
We point out that the way carriers are injected in the cavity plays a role in the relaxation dynamics. In our experiment, carriers are injected within a very tiny area, owing to resonant absorption, and a much faster time constant (about 1 ps) appears, which is related to a strong carrier diffusion at the initial stage. Based on modelling, we estimate that the volume occupied by the carriers expands from $2.5\times10^{-20}\, \mathrm{m}^3$ to $\approx8\times10^{-20}\, \mathrm{m}^3$ after few picoseconds. A very different case is when carriers are injected by an off-plane focused pump, as for instance in Ref.\cite{bose_carrier_2015}: here no diffusive contribution to the dynamics has been observed. Also, the dynamic response of the device is strongly modified if the duration of pump pulse is very short (sub-ps).
In our experiment, all the measurements have been performed with the same excitation in the Telecom spectral range and which can be generated by compact laser sources and could propagate with limited dispersion over silicon photonic circuits, therefore corresponding to more practical operating conditions.\\
We have repeated the same experiment with other cavities within the same chip, with slightly different parameters, e.g. Q factor ranging from 2000 to 3000, with some dispersion in the spectral position of the pair of resonances, which are still always spaced by approximately 10 nm. Measurements on different devices in the same fabrication batch lead to an estimate $\tau_{carr}=11.5\pm2\,\mathrm{ps}$, thereby implying a good control and reproducibility of the electrical properties of the ALD coated surface.
\section{Wavelength conversion}
Wavelength conversion consists in transferring an intensity pattern, e.g. a sequence of pulses from the pump at a specified wavelength $\lambda_p$ to a new wavelength $\lambda_s$ through all-optical modulation of a continuous wave probe at $\lambda_s$. For practical applications, also including sampling and time-demultiplexing, a meaningful repetition rate is in the GHz range. As the repetition rate increases, the average power dissipated increases too. Moreover the incomplete recovery of the gate might influence the following pulse. Therefore this measurement is a sensible benchmark for all-optical processing. The experiment is very similar to Ref. \cite{combrie_all-optical_2013} and it is performed using a mode-locked diode laser, which is tuned to the high energy resonance of the cavity by spectral slicing.\\
\begin{figure}[!t]
	\begin{center}
	\includegraphics[width=.95\columnwidth]{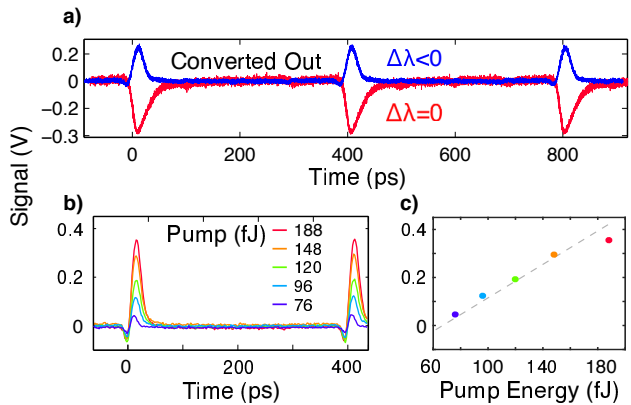}
	\caption{\label{fig:wavelength-conversion} Oscilloscope traces of the all-optically modulated CW input at $\lambda\approx1542$ nm.
	a) operation at zero signal detuning $\Delta\lambda=\lambda_s-\lambda_r=0$, and negative detuning ($\Delta\lambda<0$); b) modulated signal as a function of the pump energy and c) maximum of the signal vs. pump energy. In dashed gray line, the linear fit}
	\end{center}
\end{figure}
In our experiment, we have optimized the spectral position of the pulsed pump and the CW probe relative to the maximum achievable contrast, taking into account and partially compensating for the thermal drift. The CW probe was tuned either on resonance or was blue shifted by about 0.5 nm, which results in a downward or upward modulation of the transmitted probe (Fig.\ref{fig:wavelength-conversion}a). This is observed using a 70 GHz sampling oscilloscope (Textronik), a 40GHz photodiode (U2t) and a low-noise EDFA which amplifies the transmitted probe, while the pump is filtered out using a doubly tunable filter providing about 50 dB of rejection.\\%
The modulation amplitude is shown to  increase monotonically with pump power Fig.\ref{fig:wavelength-conversion}b,c, which indicates unsaturated operation. The maximum coupled average pump power is estimated to about 0.5 mW, which amounts to 200 fJ per pulse at a repetition rate of 2.5 GHz. The fraction of power which is absorbed (maximum 22\%, calculated) induces a temperature rise of about 3 K (the thermal resistance is $3\times 10^4\, \mathrm{K.W}^{-1}$), hence a red shift of about 150 pm.\\
Further increasing the repetition rate to 10 GHz resulted into a substantial decrease of the switching contrast which we believe to be due to the fact that a four fold increase of the dissipated power induces a large enough red shift to drive the cavity out of resonance in a microsecond time scale (related to thermal dynamics).

\section{Ultra-fast gating}
\begin{figure}[!t]
	\begin{center}
	\includegraphics[width=.95\columnwidth]{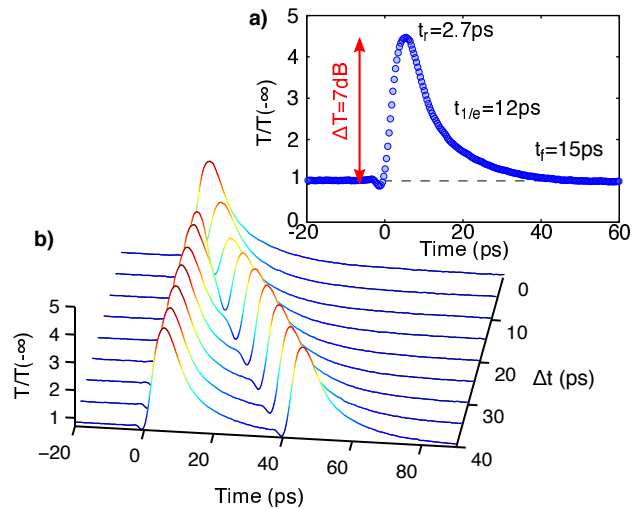}
	\caption{\label{fig:ultra-fast} Time-resolved homodyne Pump-probe measurements with $\Delta\lambda=-2.2nm$; a) relative probe transmission $T/T(-\infty)$ vs. pump-probe delay and extracted rise-time $t_r$, fall time $t_f$ and width at $1/e$, b) same as above but with a sequence of two pump pulses with distance $\Delta t$ varying from 0 ps to 40 ps.}
	\end{center}
\end{figure}
A more accurate measurement of the response of the switch is performed through a non-degenerate homodyne technique as in Fig. \ref{fig:setup}, where the spectrally broad probe is replaced with a 2 ps long pulse (approximately 0.7 nm wide), still obtained from the same femtosecond laser. We achieved a maximum switching contrast of 7 dB and a switching window at $1/e$ of 12 ps (Fig. \ref{fig:ultra-fast}a) with a pump energy of 200 fJ. Considering that the operation conditions (pump pulse power, duration and centre wavelength) where exactly the same as in the wavelength conversion experiment, this clearly shows that the response measured at the oscilloscope is affected by the instrument itself. Reducing the pump energy to about 100 fJ still allowed a contrast of 3 dB.\\
It is interesting to observe the response to a short train of pulses at very high repetition rate. In Y\"{u}ce et al.\cite{yuce_all-optical_2013}, a planar semiconductor cavity was shown to respond to a 4-pulse train at 1 THz rate. Such a fast response results from a Kerr-induced index change, owing to a specific configuration of the semiconductor alloy which suppressed the nonlinear absorption and, consequently, carriers-induced effects. However, the energy density of the focused pump is $22 \,\mathrm{pJ}.\upmu \mathrm{m}^{-2}$, implying the pump energy is in the nJ range (more than 4 orders of magnitude larger than in our case) when considering practical values for the diameter of the focused spot ($>20 \upmu \mathrm{m}$).\\.
In our experiment, the pump pulse is duplicated using a Michelson interferometer, which also allows to adjust the pulse to pulse delay $\Delta t$ accurately. When varying  $\Delta t$ from 40 ps to 0 ps, the response of the switch reveals two peaks which remain well defined (Fig. \ref{fig:ultra-fast}b) even when  $\Delta t$ decreases to $10 \mathrm{ps}$. This shows that the device is responding to a modulation at 100 GHz. We point out that this is consistent with the  maximum 140 GHz instantaneous frequency shift observed with time-resolved spectral analysis (Fig. \ref{fig:spectral_shift}b).

\vspace{-3ex}
\section{Conclusions}
An integrated fast all-optical switch needs to satisfy different requirements: energy efficiency, speed, switching contrast. Moreover, it should be operated with practical optical control signals, namely in the Telecom spectral range and with duration in the few ps range. We demonstrate the optimisation of the carrier lifetime to a close to ideal value of 12 ps. This is accomplished by passivating the surface of the photonic crystal using a conformal deposition of $Al_2O_3$ by Atomic Layer Deposition. Wavelength conversion with a time resolution of about 12 ps is demonstrated  at a repetition rate of 2.5 GHz, which is only limited by thermal effects. The response to a train of pulses shows a large-signal modulation bandwidth close to 100 GHz. The energy per pulse (on chip) is in the 100 fJ range. \\
\section{Acknowledgements}
This work was funded by the DGA (AUCTOPUSS project) and the authors would like to thank the Renatech network.

\bibliography{all-optical-gates}
\bibliographystyle{apsrev4-1}
\end{document}